\documentclass[aps,a4paper,showpacs,twocolumn]{revtex4}

\usepackage{epsfig}
\usepackage{subfigure}
\usepackage{amsmath,amssymb,color}
\usepackage[english]{babel}

\parskip=\medskipamount

\newcommand{\eq}[1]{(\ref{#1})}
\newcommand{\fig}[1]{Fig.\ref{#1}}

\newcommand{\be}{\begin{equation}}
\newcommand{\ee}{\end{equation}}

\newcommand{\barr}{\begin{array}}
\newcommand{\earr}{\end{array}}

\newcommand{\beqn}{\begin{eqnarray}}
\newcommand{\eeqn}{\end{eqnarray}}

\newcommand{\bs}{\begin{subequations}}
\newcommand{\es}{\end{subequations}}

\newcommand{\bw}{\begin{widetext}}
\newcommand{\ew}{\end{widetext}}

\newcommand\disp{\displaystyle}

\newcommand{\la}{\left<}
\newcommand{\ra}{\right>}

\newcommand{\f}{\frac}

\begin{document}

\title{From generalized directed animals to the asymmetric simple exclusion process}

\author{N. Haug$^{1,2}$, S. Nechaev$^{2,3,4}$, and M. Tamm$^{4,5}$}

\affiliation{$^1$School of Mathematical Sciences, Queen Mary University of London, London, E1 4NS,
United Kingdom \\ $^2$LPTMS, Universit\'e de Paris Sud 11, 91405 Orsay Cedex, France \\
$^3$P.N. Lebedev Physical Institute of the Russian Academy of Sciences, 119991 Moscow, Russia \\
$^5$ Department of Applied Mathematics, International Research University Higher School of
Economics, 101000, Moscow, Russia \\ $^4$ Physics Department, Lomonosov Moscow State University,
119991, Moscow, Russia}

\begin{abstract}
Using the generalized normally ordered form of words in a locally-free group of $n$ generators, we
show that in the limit $n\to\infty$, the partition function of weighted directed lattice animals on
a semi-infinite strip coincides with the partition function of stationary configurations of the
asymmetric simple exclusion process (ASEP) with arbitrary entry/escape rates through open
boundaries. We relate the features of the ASEP in the different regimes of the phase diagram to the
geometric features of the associated generalized directed animals by showing the results of
numerical simulations. In particular, we show how the presence of shocks at the first order
transition line translates into the directed animal picture. Using the evolution equation for
generalized, weighted Lukasiewicz paths, we also provide a straightforward calculation of the known
ASEP generating function.

\vspace{0.5cm}

\centerline{\it Dedicated to A.M. Vershik on the occasion of his 80th birthday}
\end{abstract}


\maketitle

\section{Introduction}

\begin{figure}[ht]
\epsfig{file=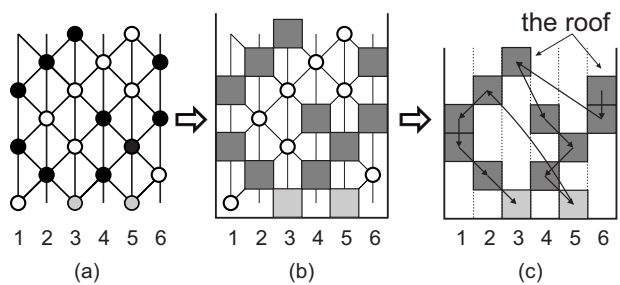, width=8cm}
\caption{(a): example of a directed animal (DA); (b)-(c): the corresponding heap of dimers.
The arrows in (c) illustrate the ``Mikado'' enumeration (see text).}
\label{an:f01}
\end{figure}

The concept of {\em heaps of pieces} was introduced by G. Viennot in 1986 \cite{Viennot86} (see
also \cite{Krattenthaler} for a review). Informally, a heap of pieces is a collection of elements
which are piled together. If two elements intersect in their horizontal projections, then the
resulting heap depends on the order in which the two are placed. In this case, the element which is
placed second is said to be \emph{above} the element placed first. On the other hand, the resulting
heap does not depend on the order in which two elements are placed if their horizontal projections
do not intersect. A special case of heaps are heaps of \emph{dimers}. The dimers can be drawn as
unit squares (boxes) which are not allowed to touch each other with their vertical edges. This
means that if we place  a box in the $k$th column and one in the $(k\pm1)$th column afterwards,
then the resulting heap is different to the one obtained from placing the boxes in the inverse
order. Figure \ref{an:f01}c shows an example of a heap of dimers.

Heaps of dimers are particularly interesting due to their relation to the model of {\em directed
animals} (DA). The term ``lattice animals'' is used as a collective name for several related models
describing the growth of aggregates, for example, molecular layers on substrates. Two-dimensional
directed animals are structures of occupied and unoccupied nodes on a lattice strip of width $n$
and infinite height. In this paper we only consider triangular lattices. The occupied sites on the
lowest row are called roots (or source points) and the DA has to satisfy the condition that each
occupied site can be reached from at least one root along a directed path containing only occupied
sites \emph{via diagonal or vertical edges} (for the triangular lattice). Figure \ref{an:f01}a shows an
example of a directed animal on such a triangular lattice and figures \ref{an:f01}b and
\ref{an:f01}c illustrate the bijection between directed animals and heaps of dimers, which works as
follows. Given a directed animal, we draw boxes around the occupied sites. This way, the DA shown
in Figure \ref{an:f01}a is redrawn as shown in Figure \ref{an:f01}b. If a box is not supported from
below, then we shift it downwards so that it is now supported. This way, we obtain Figure
\ref{an:f01}c. It can be easily seen that this mapping is in fact invertible. Namely, for a given
heap like in Figure \ref{an:f01}c, we shift upwards each box which sits on top of a box in the same
column and with it all the boxes which are above it. Then we redraw the boxes as black circles,
place white circles everywhere else and connect everything by a triangular lattice to obtain our
DA. The arrows in \fig{an:f01}c illustrate the so-called ``Mikado'' enumeration of the nodes which
will be explained below. Since we are only considering heaps of dimers in this paper, we will
shortly refer to them as heaps from now on. Also, due to the described bijection, we use the term
directed animals and heaps synonymously.

The typical problem for $N$-site DA concerns the computation of the number $\Omega(N,n|\{C\})$ of
all \emph{distinct} DA configurations in the bounding box of $n$ columns for a given configuration
of roots (base) $\{C\}$ (for example, the base of the DA in \fig{an:f01} is $\{C\}=\{3,5\}$). This
function has been computed exactly for the first time by Hakim and Nadal \cite{Hakim83} by using
algebraic methods dealing with the transfer matrix diagonalization for some spin system.

In this work, we review a different algebraic approach to directed animals which consists in
representing each DA on a lattice strip of width $n$ by an ordered word, spelled by the generators
of a locally free semi-group of $n$ generators, and show that there is a deep connection between
this group-theoretical approach and the \emph{asymmetric simple exclusion process} (ASEP) on an
open line.

\begin{figure}[ht]
\epsfig{file=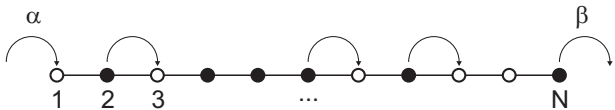, width=8cm}
\caption{the asymmetric simple exclusion process on a chain of $N=11$ sites with entering rate
$\alpha$ and escape rate $\beta$.}
\label{an:f02}
\end{figure}

The ASEP is a stochastic process on a chain of  $N$~sites which can be either occupied by a
particle or empty. A particle hops to its right with rate 1 if the right neighbouring site is
empty. For a detailed introduction and review of important results for this process, we refer the
reader to \cite{Derrida98}. The ASEP can be considered both with periodic and open boundary
conditions. In this paper, we consider open boundaries, where particles enter the chain from the
left with rate $\alpha$ and exit the chain on the right with rate $\beta$ (see \fig{an:f02}). For
this case, the probability distribution of the stationary state has been derived in
\cite{Derrida93} by a matrix ansatz. Combinatorial interpretations of the steady state weights of
the configurations have already been given in terms of pairs of paths (for $\alpha=\beta=1$)
\cite{Shapiro82}, in terms of weighted permutation tableaux \cite{Corteel07}, and in terms of
weighted binary trees \cite{Viennot07}.

In what follows we will give a new combinatorial interpretation of the stationary weights of the
ASEP on an open line in terms of directed animals. More precisely, we will demonstrate that the
partition function of the ASEP steady state on an $N$-site segment with entrance and exit rates
equal to one coincides with the partition function of $(N+1)$-site directed animals on a triangular
semi-infinite lattice strip with the topmost particle (the ``roof'') being located at the left
boundary. This correspondence can be extended towards arbitrary entrance and exit rates by defining
an appropriate weighting of the position of the leftmost root and a ``sticky" left boundary. It is
then possible to relate the features of the steady state distribution of the ASEP in the different
regimes of the phase diagram to the geometric features of the associated generalized directed
animals.


\section{Algebraic approach to directed animals}

The algebraic approach to directed animals consists in assigning to each DA--configuration an
equivalence class of words in some semi--group with special local commutation relations
corresponding to local particle configurations as shown in \fig{an:f02}. To be specific, define the
\emph{locally free semi-group}, $F_n^+$ with $n$ generators $g_1,...,g_n$, determined by the
relations
\be
g_k g_m = g_m g_k, \qquad |k-m| \ge 2.
\label{eq:01}
\ee
Each pair of neighbouring generators, $(g_k, g_{k\pm 1})$ produces a free sub-semigroup of $F_n^+$.

\begin{figure}[ht]
\epsfig{file=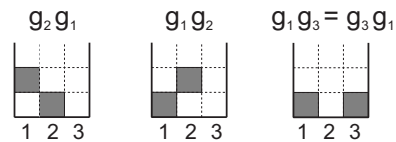,width=5cm}
\caption{Commutation relations in the group $F_n^+$ and local configurations of dimers.}
\label{an:f03}
\end{figure}

\begin{figure}[ht]
\epsfig{file=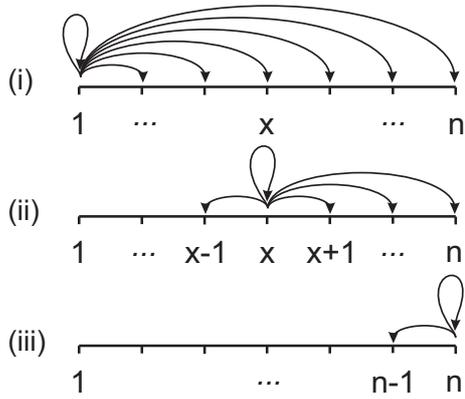, width=6cm}
\caption{The set of possible values which can be taken by the index $s_{i+1}$ if the index $s_i$ is
equal to: 1 (i), $2,...,n-1$ (ii), $n$ (iii).}
\label{an:f04}
\end{figure}

The statistical properties of locally free groups and semi-groups were investigated in detail in
\cite{Nechaev00}, where it has been shown that the partition function of an $N$--site heap in a
bounding box of size $n$ coincides with the partition function of an $N$-step \emph{Markov chain}
on $F_n^+$, or, equivalently, by the total number of equivalence classes of $N$-letter words in
$F_n^+$. Namely, to any configuration of DA one can bijectively associate an equivalence class of
words in $F_n^+$. Now each equivalence class contains exactly one word which is in \emph{normal
form}, which means that in this word, the generators with smaller indices are pushed as left as
possible in accordance with the commutation relations \eq{eq:01}. Consequently, the word
\be
W=g_{s_1} g_{s_2}\ldots g_{s_N},
\label{eq:02}
\ee
is in ordered form if and only if the indices
$s_1,...,s_N$ satisfy the following conditions, graphically represented in \fig{an:f04}.

\begin{itemize}
\item[(i)] If $s_i=1$ then $s_{i+1}\in\{1, 2,..., n\}$;
\item[(ii)] If $s_i=x$ ($2\le x\le n-1$)\\
then $s_{i+1}\in\{x-1, x, x+1,...,n\}$;
\item[(iii)] If $s_i=n$ then $s_{i+1}\in\{n-1, n\}$.
\end{itemize}

Thus, any $N$-site heap in a bounding box of $n$ columns can be uniquely represented by
an $N$-letter ordered word, ``spelled'' by the generators of $F_n^+$. For example, the normally
ordered word
\be
W=g_3\; g_2\; g_1\; g_1\; g_2\; g_5\; g_4\; g_5\; g_4\; g_3\; g_6\; g_6
\label{eq:03}
\ee
uniquely represents the 12-site directed animal shown in \fig{an:f01}a.

For a given heap, the corresponding normally ordered word can be obtained by an algorithm which
sets a constructive geometrical way of normal ordering. We call this enumeration procedure the
``Mikado ordering" since it resembles the famous Mikado game, the goal of which consists in the
sequential removal of the boxes from a random pile, one-by-one, without disturbing the other
elements.  To proceed, define in a heap a set of top sites, each of which can be removed from the
heap without disturbing the rest of the pile. We call these elements the ``roof", ${\cal T}$, of
the heap. Remove the rightmost element of ${\cal T}$. In the updated roof, ${\cal T}'$, remove
again the rightmost element to get ${\cal T}''$, and so on, until the heap is empty. The sequence
of one-by-one removed elements is normally ordered and uniquely enumerates the heap (i.e. the
directed lattice animal). This fact is established in Lemma 3 of \cite{Nechaev00}. For the heap
shown in \fig{an:f01}c, the Mikado ordering is depicted by the sequence of arrows and coincides
with \eq{eq:03} (note that the topmost element in the 4th column does not belong to the roof as it
cannot be removed without disturbing the topmost element in the 3rd column, which is above it).

\section{Matrix ansatz for generalized DA and ASEP}

We now introduce the partition function $\Omega_{i,j}(N+1,n)$, which enumerates all the
$(N+1)$-particle heaps in the bounding box of $n$ columns whose Mikado ordering has its first
element in the $i$-th and its last element in the $j$-th column. This function can be expressed in
terms of a local $(n\times n)$ transfer matrix, $M$, with transitions described by the rules
(i)-(iii) (see also \fig{an:f04}), namely
\be
\Omega_{i,j}(N+1,n) = \la v_{i} | M^N | v_{j} \ra,
\label{eq:04}
\ee
where $\la v_{k} \right| =(\overbrace{0,...,0,1,0...0}^{\rm n})$ with a one in the $k$-th position,
and, as usual, $\left| v_{k} \ra=\la v_{k} \right|^{\top}$. For reasons which will become clear in
the following, we are mostly interested in the values of $\Omega_{i,j=1}(N+1,n)$. It is instructive
to introduce the generating function
\be
Z_{N+1}(n,\alpha)=\sum_{i=1}^n \Omega_{i,1}(N+1,n) \alpha^{1-i} = \la v_{in} | M^N | v_{1} \ra,
\ee
where $\la v_{in} \right| =(\overbrace{1,\alpha^{-1},\alpha^{-2},\alpha^{-3}...}^{n})$. Now the
transfer matrix $M$ allows a natural decomposition in ``forward'' ($D$) and ``backward'' ($E$)
parts, associated with arbitrarily far jumps to the right and one-step jumps to the left (see
\fig{an:f04}). Namely, we can write $M=D+E$, where
\be
D= \left(\begin{array}{ccccc}
1 & 1 & 1 & \ldots & 1 \\
0 & 1 & 1 & \ldots & 1 \\
0 & 0 & 1 & \ldots & 1 \\
\vdots & \vdots & \ddots & \ddots & \vdots \\
0 & 0 & \cdots & 0 & 1
\end{array}\right); \quad
E= \left(\begin{array}{ccccc}
0 & 0 & 0 & \ldots & 0 \\
1 & 0 & 0 & \ldots & 0 \\
0 & 1 & 0 & \ldots & 0 \\
\vdots & \vdots & \ddots & \ddots & \vdots \\
0 & 0 & \cdots & 1 & 0
\end{array}\right).
\label{eq:05}
\ee

There is a striking similarity between this result and the celebrated exact solution of the
asymmetric simple exclusion process (ASEP) on a line \cite{Derrida93}. Let us briefly recall this
well-known result. The steady state of this process can, according to \cite{Derrida93}, be
calculated via the following procedure know as the ``matrix ansatz". Introduce two formal operators
$\widetilde{D}$ and $\widetilde{E}$ which satisfy
\be
\widetilde{D} + \widetilde{E} = \widetilde{D}\widetilde{E}
\label{eq:ans1}
\ee
and two vectors $\la \widetilde{v}_{\rm in}\right|$ and $\left|\widetilde{v}_{\rm out}\ra$, such that
\be
\la \widetilde{v}_{\rm in} \right|\widetilde{E} = \alpha^{-1} \la \widetilde{v}_{\rm in}\right|~;~
\widetilde{D}\left|\widetilde{v}_{\rm out}\ra =
\beta^{-1} \left|\widetilde{v}_{\rm out}\ra.
\label{eq:ans2}
\ee

Now one can show that the probability of observing any given ASEP configuration in the steady state
is proportional to a matrix element of the type $\la \widetilde{v}_{\rm in}|...|\widetilde{v}_{\rm
out}\ra$, where for the dots one should insert a sequence of the operators $\widetilde{D}$ and
$\widetilde{E}$, with $\widetilde{D}$ and $\widetilde{E}$ corresponding to occupied and empty
sites, respectively. For example, the probability of the configuration shown in \fig{an:f02} is
proportional to $\la \widetilde{v}_{\rm
in}\right|\widetilde{E}\widetilde{D}\widetilde{E}\widetilde{D}
\widetilde{D}\widetilde{D}\widetilde{E}
\widetilde{D}\widetilde{E}\widetilde{E}\widetilde{D}\left|\widetilde{v}_{\rm out}\ra$. The sum
$\widetilde{Z}_N(\alpha,\beta)$ of matrix elements over all possible configurations, which plays a
role very similar to the partition function of the steady-state ASEP, can be written as
\begin{multline}
\widetilde{Z}_N(\alpha,\beta) = \la {\widetilde{v}}_{\rm
in}|(\widetilde{E}+\widetilde{D})^N|{\widetilde{v}}_{\rm out}\ra = \la {\widetilde{v}}_{\rm
in}|(\widetilde{M})^N|{\widetilde{v}}_{\rm out}\ra.
\label{eq:07}
\end{multline}

For arbitrary $\alpha$ and $\beta$, the algebra defined by \eq{eq:ans1} and \eq{eq:ans2} has no
finite-dimensional representations. However, there exist many infinite-dimensional representations,
among which the most interesting for us is constructed as follows. Set $\la \widetilde{v}_{\rm
in}\right|=(1,\alpha^{-1},\alpha^{-2}, \alpha^{-3},\dots)$, $\la \widetilde{v}_{\rm out}
\right|=(1,0,0,\dots)$ and choose the matrices $\widetilde{D}$ and $\widetilde{E}$ as
\be
\widetilde{D}= \left(\begin{array}{cccc}
\frac{1}{\beta} & \frac{1}{\beta} & \frac{1}{\beta} & ...  \\
0 & 1 & 1 & \ldots  \\
0 & 0 & 1 & \ldots  \\
\vdots & \vdots &  & \ddots
\end{array}\right); \quad
\widetilde{E}= \left(\begin{array}{cccc}
0 & 0 & 0 & \ldots  \\
1 & 0 & 0 & \ldots  \\
0 & 1 & 0 & \ldots  \\
\vdots & \vdots &  & \ddots
\end{array}\right).
\label{eq:08}
\ee

It is easy to check that conditions \eq{eq:ans1} and \eq{eq:ans2} are satisfied. Furthermore, the
similarity between \eq{eq:05} and \eq{eq:08} is strikingly clear. Indeed, we immediately get
\be
\widetilde{Z}_N(\alpha,\beta=1)= \lim_{n\to \infty} Z_{N+1}(n,\alpha).
\label{eq:08a}
\ee

This explicitly demonstrates that in the limit $n\to\infty$, the generating function of
$N+1$-particle heaps with a topmost particle in the first column and activity $\alpha$ associated
to the position of the first particle (which is the leftmost particle in the lowest row), coincides
with the ``partition function'' of the stationary ASEP chain of $N$ sites in the case $\beta=1$.

\begin{figure}[ht]
\epsfig{file=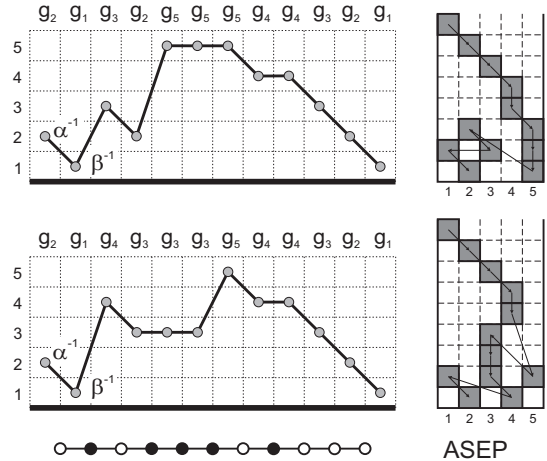, width=7cm}
\caption{Two generalized directed animals corresponding to the same ASEP configuration (below) and
the Markov chains representing the associated ordered words. The horizontal coordinate stands for
the index of a letter in the word. All jumps starting from position $x=1$ carry the weight
$\beta^{-1}$ and a first letter $g_x$ contributes the weight $\alpha^{-(x-1)}$. The last letter is
always $g_1$.}
\label{an:f05}
\end{figure}

The correspondence between DA and the ASEP for arbitrary values of $\alpha$ \emph{and} $\beta$ is
established as follows. Each ASEP configuration corresponds to a {\emph set} of \emph{pyramids}
\cite{Viennot86}, i.e. heaps with a roof consisting of a single element in column 1. Now we can
associate the sequence of ``backward'' and ``forward'' jumps with the ASEP configuration such that
backwards jumps correspond to a hole in the ASEP sequence, and forward jumps to a particle -- see
\fig{an:f04} (here ``no jump" is considered as a ``forward jump" with length zero). Since there can
be forward jumps of different lengths, there are, generally speaking, many different DAs
corresponding to a single ASEP configuration. This fact is depicted in \fig{an:f05}, where we show
two heaps of dimers corresponding to the same ASEP configuration.

Now the weight of a given ASEP configuration in the steady state on a line is proportional to the
sum of the weights of \emph{all} corresponding heaps, where the weight of a given heap equals
$\alpha^{1-x} \beta^{1-y}$ with $x$ being the coordinate of the column in which the leftmost root
is located, and $y$ being the number of elements of the heap in column 1. Using the Mikado
enumeration, one can rephrase this statement as follows: (a) the first letter in the normally
ordered word associated to a specific DA has weight $\alpha^{1-x}$, (b) each generator $g_1$
(except for the last letter) carries the weight $\beta^{-1}$, while all the other generators have
weight $1$, (c) the last letter in the normally ordered word is always $g_1$, (d) to get a weight
of an ASEP configuration one has to sum over all corresponding DAs in which any pair $g_i g_k$ with
$k\ge i$ ($(k,i)\in \{1,\dots,n\}^2$) corresponds to a particle, while a pair $g_i g_{i-1}$ ($i\in
\{2,\dots,n\}$) corresponds to a hole. In Table \ref{tab:01}, we summarize the correspondence
between the stationary ASEP and DA. We note several important facts about this correspondence.

\begin{table}[ht]
\epsfig{file=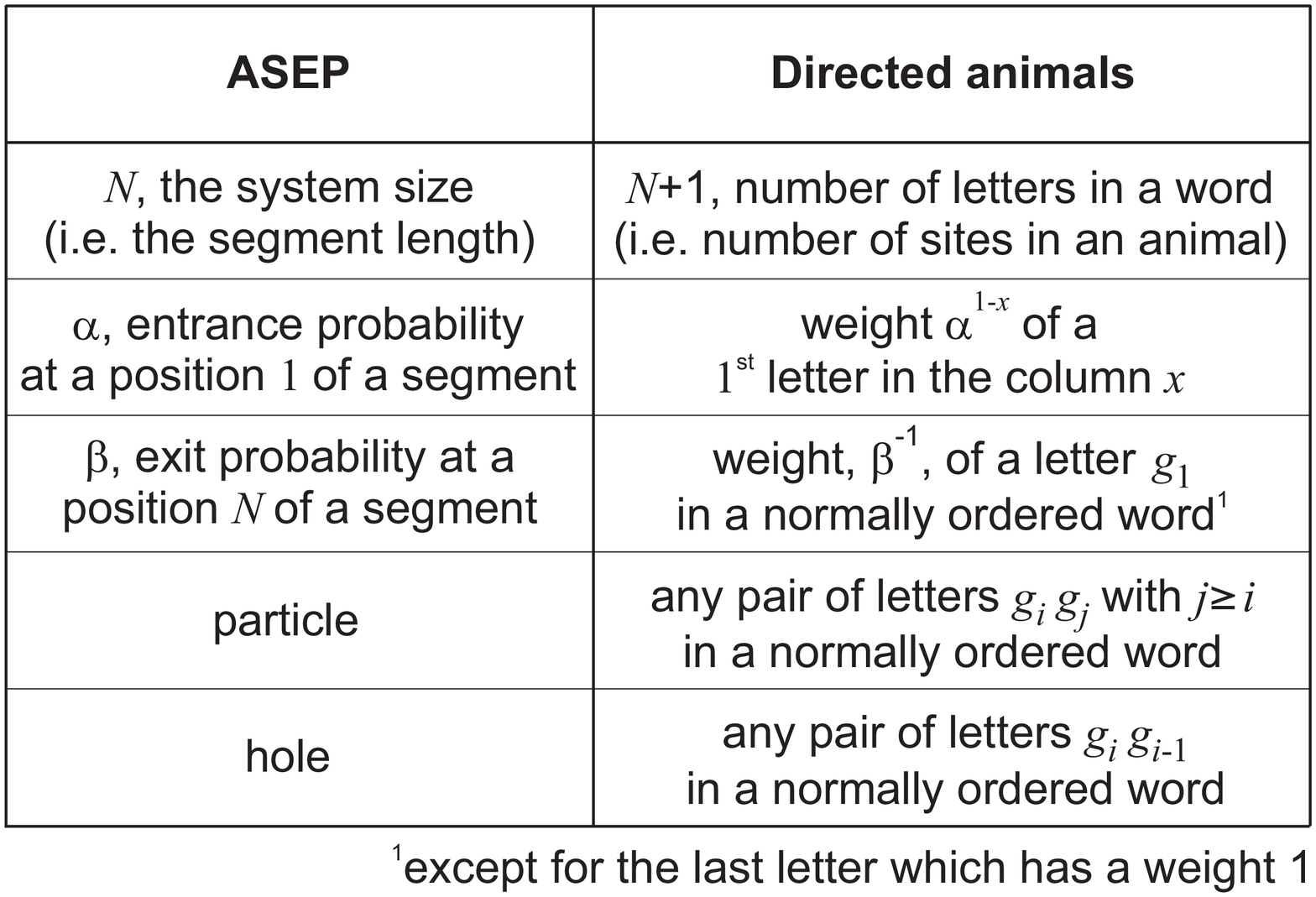, width=8cm}
\caption{Correspondence between the asymmetric simple exclusion process and directed animals on
strip of width $n$ in the limit $n\to\infty$.}
\label{tab:01}
\end{table}

First, there is no straightforward analog of time in the heap picture, and there is no time
evolution imposed on the DAs. From the mapping revealed above one immediately gets the\,\emph{
steady state} probabilities of the ASEP configurations, but not the underlying time evolution.

Second, formally, the mapping is exact only for $n\to\infty$ (i.e., for directed animals on a
quarter-plane with no boundary on the right. However, the condition that the last particle should
be put in column 1 dictates that all $N$-step trajectories stay in columns $x\leq N$, so $n
\geq N$ is enough to make the mapping exact.

Third, the original ASEP problem has a well-known particle-hole symmetry, i.e. if one replaces the
particles with holes and vice versa, reverses the direction of the flow and interchanges $\alpha
\leftrightarrow \beta$, one returns to the original problem. This symmetry is evident in the formal
algebraic matrix ansatz, but broken down by the representation \eq{eq:08}, which makes it a
bit artificial in the original ASEP model. The interpretation in terms of Mikado-ordered DAs gives,
however, a natural, intuitive meaning to the representation \eq{eq:08}. The connection between the
DA and the ASEP problems also shows that there is actually a hidden symmetry in the DA model,
namely a symmetry between the \emph{position} of the leftmost root and the \emph{number of visits}
of the column $x=1$. In other words, the partition function of $N$-particle DAs with a single roof
particle in the first column, $s$ particles with weight $\beta^{-1}$ in the first column and
leftmost root in column $k$ with weight $\alpha^{1-k}$, coincides with the partition function of
$N$-particle DA with a single roof particle in the first column, $k$ particles in the first column
with weight $\alpha$ in total and leftmost root in column $s$ with weight $\beta^{1-s}$.

Now we can better understand the statistics of standard, non-weighted directed animals
($\alpha=\beta=1$). The point $(\alpha,\beta)=(1,1)$ lies deeply in the maximum-current phase of
the ASEP (\cite{Derrida98}, see also the next section). This means that the corresponding ASEP
steady state is dominated by configurations where the particle density is equal to $1/2$. In terms
of heaps, this means that for $N$ large enough, there are typically equal numbers of left and right
jumps in the Mikado ordering (compare with the numeric results shown in the next section). This
might seem counter-intuitive, as rightward jumps can have an arbitrarily big length while leftward
jumps always have length $1$. Note, though, that ``rightward" jumps can also have length $0$, so
the average length of a rightward jump can well turn out to be $1$.

Small changes in $\alpha$ or $\beta$ do not move the ASEP out of the maximum-current phase, and
thus the concentrations of forward and backward jumps stay equal also for $\alpha$ and $\beta$
slightly differing from $1$. One has to go as far as $\alpha=1/2$ or $\beta=1/2$ to see a big
change in the behavior of the heaps. We suggest the reader to compare this result to the
adsorption-desorption transition of a random walk (polymer chain) on a half-line with a potential
well at $x=0$, where the change in typical trajectories also occurs for the potential well depth
$\beta^{-1}=2$ \cite{Naidenov01}.

\section{Simulation of generalized heaps}

\begin{figure}[ht]
\epsfig{file=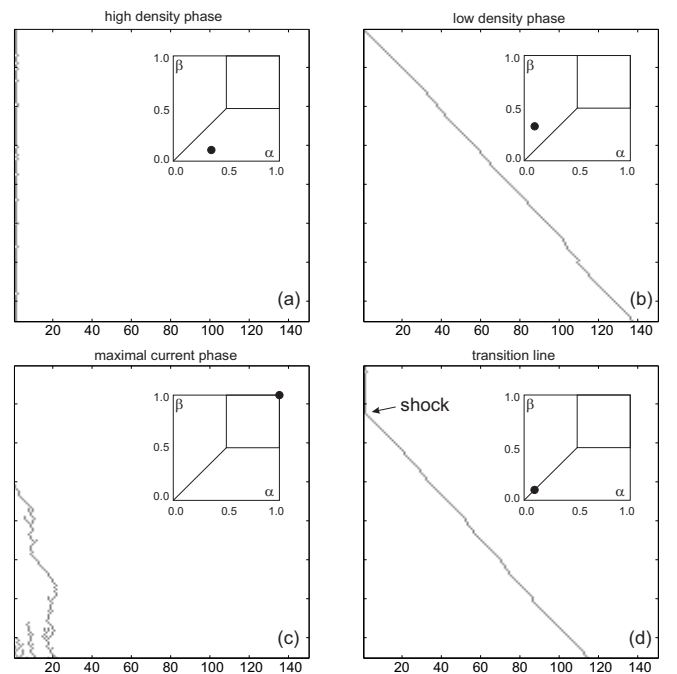,width=8.5cm}
\caption{Numerically generated random generalized heaps of $N$=150 dimers for different values of
$\alpha$ and $\beta$. (a): $\alpha = 0.3$, $\beta = 0.1$, (b): $\alpha = 0.1$, $\beta = 0.3$, (c):
$\alpha = \beta = 1$, (d): $\alpha = \beta = 0.1$. The inserts in each picture show the position in
the phase diagram of the ASEP, the black lines mark the phase boundaries.}
\label{an:f07}
\end{figure}

For different values of $\alpha$ and $\beta$, we have numerically generated corresponding
generalized heaps of $N=150$ particles. The simulation was carried out by generating random
$N=150$-step generalized Lukasiewicz paths with fixed endpoint at $x=1$ \cite{Lehner03} and weighting
of the steps and initial position according to \fig{an:f05}. The algorithm we used is described in
\cite{Kamenetskii81}. \fig{an:f07} shows the resulting pictures.

In the high density phase $\beta < \alpha < \f 12$, the typical heaps are roughly vertical piles in
the first column with only a few boxes sticking out into the second column. The number of boxes
which are supported from below right, corresponding to a hole in the associated ASEP configuration,
is very small, the particle density of the corresponding ASEP configuration is close to 1 in this
case -- see \fig{an:f07}a.

In the low density phase $\alpha < \beta < \f 12$, the typical heap roughly follows a diagonal
line, going from below right to the top left. This corresponds to a very low particle density of
the corresponding ASEP configuration -- see \fig{an:f07}b.\\
For both $\alpha$ and $\beta$ greater than $\f 12$, one obtains less regular pictures with, on
average, as many boxes which are supported from below left or sit on top of another box as boxes
supported from below right. This means that the corresponding ASEP configuration has a particle
density close to $\f 12$. The maximal current case $\alpha=\beta=1$ is depicted in \fig{an:f07}c.\\
At the first-order transition line $\alpha = \beta < \f 12$, one observes heaps which roughly
consist of a diagonal line below, followed by a straight vertical pile in the first columnn. This
means that the corresponding ASEP configuration is divided into a region with very low density on
the left and a region with very high density on the right. However, the size of the two regimes
varies. The point in the ASEP chain at which the two density regimes meet each other is identified
as a \emph{shock} -- see \fig{an:f07}d.

\section{Stationary ASEP as polymer wetting}

Let us sketch the derivation of the stationary ASEP partition function (\ref{eq:07}). Although the
answer is well known since the pioneering works \cite{Derrida93} and has been derived with
different nuances in some subsequent works (see, for example, \cite{Blythe07,Depken04}), we would like
to emphasize the deep analogy of the ASEP generating function with the generating function of the
\emph{wetting problem} on a one-dimensional adsorbing substrate \cite{Naidenov01,Gangardt07}. In a
general setting, wetting implies the interface pinning by an impenetrable solid. Problems of
interface statistics in the presence of a hard wall were addressed in many publications (see, for
example, \cite{Abraham80,Abraham86} and references therein). The most interesting question concerns the
nature of the wetting or pinning-depinning transition of the interface controlled by parameters of
its interactions with the substrate. To the best of our knowledge, the similarity of the analytic
structures of the generating functions for asymmetric exclusion and wetting has been briefly touched
only in the review \cite{Blythe07}. The connection between the ASEP and pinned interface statistics allows
us, as we have seen in the previous section, to get a simple and transparent view on the nature
of shocks. Conversely, this connection raises open questions whether the fluctuations of the
interface density in vicinity of the pinning-depinning transition could exhibit the KPZ scaling
seen near the ASEP shock profiles as pointed out in \cite{Janowsky92}.

Define $Z_N(x,\alpha,\beta)$, the partition function of the $N$-step trajectories on the
semi-infinite discrete line ($n\to\infty$) with allowed steps and weighting as shown in
\fig{an:f04}, and with final position in $x$. For shortness we write $Z_N(x,\alpha,\beta) \equiv Z_N(x)$. This function can be expressed in terms of a matrix product as 
\be 
Z_N(x) = \la v_{\rm in} \right| \widetilde{M}^{N-1} \left| v_x \ra.
\label{matrixproduct}
\ee
 
The quantity of our interest is $Z_N(x=1)$. From \eqref{matrixproduct}, we obtain the
following recursion relation, valid for any $N\ge 0$ (compare to
\cite{Derrida93}):
\be
\left\{\begin{array}{rlll}
Z_{N+1}(x) & = & \disp \beta^{-1}Z_N(1)+\sum_{y=2}^{x+1} Z_N(y) & x = 1,2,..., \\
Z_{N=0}(x) & = & \disp \alpha^{x-1} & x = 1,2,..., \medskip \\
Z_N(x) & = & 0 & x=0.
\end{array}
\label{eq:09}
\right.
\ee
Introduce the generating function
$$
W(s,x) = \sum_{N=0}^\infty Z_N(x) s^N;\; Z_N(x)=\frac{1}{2\pi i}\oint\limits_{C}
\frac{W(s,x)}{s^{N+1}} ds
$$
with a suitably chosen closed contour $C$ around the origin. In what follows we denote $W(s,x)
\equiv W(x)$ for shortness.

Defining now $Q(x)=s^{x/2}W(x)$ and using the Kronecker $\delta$--symbol, where $\delta_{x,1}=1$
for $x=1$, and 0 otherwise, we can rewrite \eq{eq:09} as a single equation in a symmetrized form,
which has straightforward interpretation in terms of the wetting generating function \cite{Gangardt07}
on a semi-infinite line $x\ge 0$. We get
\begin{multline}
\Big\{Q(x)-\sqrt{s}\big(Q(x-1)+Q(x+1)\big) \medskip \\
-s^{x/2}\alpha^{1-x}(1-\alpha)\Big\}(1-\delta_{x,1}) \medskip \\
+\left\{Q(x)\f{\beta-s}{\beta}-\sqrt{s}Q(x+1)-s^{x/2}\right\}\delta_{x,1} = 0.
\label{eq:13}
\end{multline}
Applying the Fourier transform
$$
\mathcal{Q}(q) = \sum_{x=0}^\infty Q(x) \sin q x; \quad Q(x) =
\f{2}{\pi}\int_{0}^\pi \mathcal{Q}(q) \sin q x\, dq
$$
to equation \eq{eq:13}, we obtain
\begin{multline}
(1-2\sqrt{s}\cos q)\mathcal{Q}(q)-\f{s}{\beta}\sin q\, Q(1)- \medskip \\
-(\alpha-\alpha^2) \sum_{x=2}^\infty\left(\f{\sqrt{s}}{\alpha}\right)^x \sin q x\,-\sqrt{s}\sin q=0.
\label{eq:18}
\end{multline}
The solution for $\mathcal{Q}(q)$ reads
\be
\mathcal{Q}(q) = \f{\f{s}{\beta}\sin q\, Q(1)+f(q)}{1-2\sqrt{s}\cos q}.
\label{eq:19}
\ee
where we have defined $f(q)$ as
\be
\begin{array}{lcl}
f(q)&=&\disp(\alpha-\alpha^2)\sum_{x=2}^\infty \left(\f{\sqrt{s}}{\alpha}\right)^x
\sin q x+\sqrt{s}\,\sin q,\medskip\\
\disp &=&\disp\f{\alpha^2(1-2\sqrt{s}\cos q)+s\alpha} {\alpha^2-2\sqrt{s}\alpha\cos
q\,+s}\sqrt{s}\sin q,\medskip
\end{array}
\label{eq:14a}
\ee
Remembering that $W(1)=s^{-1/2}Q(1)$, inserting the expression \eq{eq:14a} for $f(q)$ into
\eq{eq:19} and applying the inverse Fourier transform, we end up with
\be
\begin{array}{lcl}
W(s,1) & = & \f{\disp \f{\disp 2}{\pi\sqrt{s}} \int_{0}^\pi \f{f(q) \sin q}{1-2\sqrt{s}\cos
q}dq}{\disp 1-\f{2s}{\pi\beta}\int_{0}^\pi \f{\sin^2 q}{1-2\sqrt{s}\cos q}dq}\medskip\\
& = & \disp \f{4\alpha\beta}{\left(2\alpha-1 + \sqrt{1-4s}\right)\left(2\beta-1 +
\sqrt{1-4s}\right)}
\end{array}
\label{eq:20}
\ee
as an explicit expression for the generating function of the stationary ASEP partition function,
$Z_{N}(x=1,\alpha,\beta)$. Note that the roots in the denominator are positive for $s < \f 14$. For
$\alpha$ and $\beta<\f 12$, the generating function \eq{eq:20} has two pole singularities at
\be
s_1=\alpha(1-\alpha),\quad s_2=\beta(1-\beta),
\label{eq:21}
\ee
which are both smaller or equal to $\f 14$. For $\alpha$ and $\beta > \f 12$, these poles leave
the real axis and the branching point $s_3 = \f 14$ becomes the dominant singularity. Depending on
which singularity is dominant, one recovers the known phase diagram of the ASEP, with the three
borders $\alpha = \beta < \f 12$, $\alpha = \f 12$ and $\beta = \f 12$ \cite{Derrida98}.

As one sees, both generating functions, of the ASEP and of the wetting problem, have similar
analytic structures; they diverge at branching points, which signals the existence of a phase
transition. However, the behavior of the function $W(s,1)$ is far more rich: it has two possible
singularities $s_1$ and $s_2$ controlled by \textit{two} independent parameters, $\alpha$ and
$\beta$. Thus, in the thermodynamic limit $N\to\infty$, the ASEP ``free energy'', $f(\alpha,\beta)$
strongly depends on the parameters $\alpha$ and $\beta$ and is determined by the singularity which
is closest to zero :
\be
f(\alpha,\beta)=-\ln \min\{s_1(\alpha), s_2(\beta),s_3\}.
\label{eq:23}
\ee

\section{Summary}

In this letter we have established the connection between generalized directed animals on a
semi-infinite strip with adsorbing boundary and special initial particle distribution with the
stationary state configurations of the asymmetric simple exclusion process. Given the relation
between directed animals and the ASEP, we analysed how the features of one model translate into
features of the other one. We simulated generalized directed animals (heaps respectively) in the
different regimes of the ASEP phase diagram and discussed the shape of the typical pictures
obtained. In particular, we were able to observe shock configurations at the first order transition
line between the low and the high density phase of the ASEP. We also noted a hidden symmetry of the
directed animals model by making use of the known particle-hole symmetry in the ASEP.\\
\indent The random walk picture of directed animals which resulted from the normal order
representation of directed animal configurations (associated with the locally free group), allowed
us to regard the stationary ASEP as a sort of wetting model on a one-dimensional adsorbing
substrate. Using the evolution equation for this random walk, we provided a simple derivation of
the ASEP generating function on a one-dimensional line.

The authors are grateful to A. Vershik for numerous discussion of the problem. N.H. and M.T. would
like to thank the LPTMS for the warm hospitality. This work was partially supported by the grants
ANR-2011-BS04-013-01 WALKMAT, FP7-PEOPLE-2010-IRSES 269139 DCP-PhysBio, as well as by a MIT-France
Seed fund and the Higher School of Economics program for Basic Research.

\bibliography{mybib}{}
\bibliographystyle{unsrt} 
\end{document}